\documentstyle[12pt]{article}
\begin{document}
\thispagestyle{empty}
\begin{flushright}
UA/NPPS-20-00
\end{flushright}
\vspace{3cm}

\begin{center}
{\large{\bf Polyakov's spin factor and new algorithms
     for efficient perturbative computations in QCD}\\}

\vspace{1cm}

A. I. Karanikas and C. N. Ktorides\\
\smallskip
{\it University of Athens\\
Department of Physics\\
Nuclear and Particle Physics Section\\
Panepistimiopolis\\
GR-15771 Athens, Greece\\}
\end{center}

\vspace{2cm}

\begin{abstract} 
Polyakov's spin factor enters as a weight in the path-integral description
of particle-like modes propagating in Euclidean space-times, accounting
for particle spin. The Fock-Feynman-Schwinger path integral applied to QCD
accomodates Polyakov's spin factor in a natural manner while, at the same
time, it identifies Wilson line (loop) operators as sole agents of 
interaction dynamics among matter and gauge field quanta. A direct
application of such a separation between spin content and dynamics is the
emergence of master expressions for the perturbative series involving
either open or closed fermionic lines which provide new,
comprehensive approaches to perturbative QCD.

\end{abstract}

\newpage
\setcounter{page}{1}
{\bf 1. Introductory remarks.} The Fock[1]-Feynman[2]-Schwinger[3] path
integral, also known as worldline
formalism [4-7], constitutes a tool by which one
is able to study a quantum field theoretical system in terms of the
propagation of its particle quanta. Polyakov's path integral [8], on the
other
hand, amounts to a direct formulation regarding the (Euclidean) space-time 
propagation of particle-like entities. Its novel ingredient is the
so-called {\it spin factor} which enters the path integral as a weight 
accounting for the spin of a particle that traverses a given closed
contour\footnote{Polyakov actually addressed himself to the case of spin-1/2
entities. Subsequently, Korchemsky and Korchemskaya [9] discussed
extensions
to higher spins.}. Polyakov's scheme is distinquished by its geometrical
mode of description and serves, via its extension to two-dimensional 
worldsheets, as a prototype for discussing the propagation of quantum strings.

Inspired by Polyakov's work we have been able to reformulate [10] the
Fock-Feynman-Schwinger path integral for gauge field systems with spin-1/2
matter fields so that the spin factor, associated with the propagation of
the latter, explicitly makes its entrance. At the same time, the
dynamics operating on the matter particles is carried by Wilson line
(loop) operators.

Let us display relevant expressions corresponding to the `effective'
action functional ${\cal W}[A]$ and the full fermionic propagator
$G(x,y|A)$ in the presence of background gauge fields. They read,
respectively, as follows:

\begin{eqnarray}
{\cal W}[A]=-\int_0^\infty\, \frac{dT}{T}e^{-Tm^2}\int_{x(0)=x(T)}
& &{\cal D}x(\tau)e^{-\frac{1}{4}\int_0^T d\tau \dot{x}^2(\tau)}tr
Pexp\left\{\frac{i}{4}\int_0^T
d\tau\,\sigma\cdot\omega\right\}\nonumber\\ &&\quad\times Tr_cPexp\left[ig
\int_0^Td\tau\dot{x}\cdot A(x(\tau))\right],
\end{eqnarray}
where $tr(Tr_c)$ denotes trace over $\gamma$-matrix (color)and $P$ stands
for path ordering, and

\begin{eqnarray}
iG(x,y|A)&=&\int_0^\infty dTe^{-Tm^2}\int_{\stackrel{x(0)=x}{x(T)=y}}
{\cal D}x(\tau)
[m-{1\over 2}\gamma\cdot\dot{x}(T)]
exp\left[-{1\over 4}\int_0^T d\tau {\dot{x}^2(\tau)}\right]\nonumber\\
& &\times  P\,exp\left[{i\over 4}\int_0^Td\tau\,\sigma\cdot
\omega\right] P\,exp\left[ig\int_0^Td\tau\dot{x}(\tau)\cdot
A(x(\tau))\right].
\end{eqnarray}
In the above formulae colour indices have been supressed as $A_\mu\equiv
A_\mu^aT^a$, with the generators $T^a$ in the fundamental representation,
while $\sigma\cdot \omega$ actually stands for
$\sigma_{\mu\nu} \omega_{\mu\nu}$, with $\sigma_{\mu\nu}\equiv
{1\over 2i}[\gamma_\mu,\gamma_\nu]$ and

\begin{equation}
\omega_{\mu\nu}(x)={T\over
2}(\ddot{x}_\mu\dot{x}_\nu-\dot{x}_\mu\ddot{x}_\nu).
\end{equation}  
We normalize our expressions so that, for vanishing gauge potentials,
${\cal W}[A]$ is unity and $G(x,y|A)$ coincides with the free fermionic
propagator.

We shall refer to the quantity
\begin{equation}
\Phi(C)\equiv Pexp\left\{\frac{i}{4}\int_0^T
d\tau\,\sigma\cdot\omega\right\},
\end{equation}
with $C$ some path path parametrized by $\tau\in [0,T]$, as spin factor
expression or, simply, spin factor, even though the latter
characterization
strictly holds for a {\it closed} path, as in (1), and is given by 
$tr\Phi(C)$. It serves to properly account for the reentry
of the contour by a spin-1/2 matter particle through the involvement of
the geometrical entity of torsion. The exact same expression enters the
Green's function, only now it pertains to an open path $L$. As it turns
out [11], its impact is exclusively associated with points at which a
four-momentum is conveyed to the fermionic contour through the emission
or absorption of a gauge field quantum.

As already mentioned, the sole agent of the dynamics which involves the
matter particles is the Wilson line (loop) operator whose corresponding
contours coincide with the spin-1/2 particle paths. One might view the
overall situation as if the $\bar{\psi}\gamma_\mu\psi A^\mu$ term in the
field Lagrangian has been re-organized into a geometrical and a dynamical
factor operating independently (albeit at the same points) on the
aformentioned paths. The full dynamical treatment of gauge fields  
implicates the corresponding functional integration weighted by the
Yang-Mills action along with whatever else it entails (gauge fixing terms,
etc.).

In what follows we shall apply the above formalism to perturbative QCD
illustrating the role played by the spin factor in arriving at general
expressions which provide alternative ways of forming perturbative
expansions. Two different viewpoints will be taken. The first, to be
considered in the next section, entails carrying the path integral from
the very beginning with the spin factor surviving as a momentum dependent
quantity. In this way one is able to bring each term of the perturbative 
expansion into a form that
can be directly compared with the Feynman diagrammatic logic, albeit
through a different and hopefully more economical organization. A relevant
analytic computation to low order,
{\it will} be conducted, cf. section 3, nevertheless the
real usefulness of our results can be fully assessed  once numerical
methodologies are applied towards the evaluation of higher order
perturbative contributions. A
second possibility offered by our approach leads to a perturbative
expansion which is characterized by its space-time perspective. 
The basic computational task, in this case, reduces to
determining correlators with respect to a particle-like, i.e. one
dimensional, action over bosonic and fermionic variables into which the
spin factor is incorporated. The relevant
analysis will be presented in  section 4.

\vspace{.2 cm}

{\bf 2. Master expressions for Feynman-type perturbative expansions.}
Relations (1) and (2), as well as similar ones for higher order
Green's functions, amount to a recasting of the given quantum
field theoretical system by which nothing of its original content has
been lost\footnote{This is so because terms in the field theoretical
action that have been integrated over are the ones involving the spin-1/2 
fields, i.e. they are of Gaussian form.}. The question of interest, then,
is whether
the Fock-Feynman-Schwinger reformulation of the quantum field theory, QCD
in particular, presents advantages at a {\it practical} level. As already
indicated, our focus,
in this paper, will lie on perturbation theory algorithms. For closed
fermionic contours it is already known, from the work of Strassler [6],
that the worldline framework reproduces the string theory-inspired rules 
of Bern and Kosower [12] which promote efficient perturbative computations
in QCD. For open contours, on the other hand, the situation is somewhat
more demanding on account of qualitatively different boundary conditions
that one needs to accomodate. It is to this case that we shall direct our
attention throughout this paper, restricting ourselves to situations where
a single fermionic line is involved. A point of reference is furnished
by expression (2) for the quark propagator in a background of gluonic
fields, even though we shall eventually generalize our considerations so
as to include gluons as dynamical entities in both virtual and
external states. The specific methodology that will characterize our work
in this section aims at direct comparisons with the Feynman diagrammatic
logic. 

Expanding the Wilson (line) exponential in (2),
we readily obtain a series of parametric integrations of the form
\begin{equation}
(ig)^M\int_0^Td\tau_M...\int_0^Td\tau_1\theta(\tau_M,...,\tau_1)
A(x(\tau_M))\cdot\dot{x}(\tau_M)...A(x(\tau_1))\cdot\dot{x}(\tau_1)
\end{equation}
which find themselves immersed into larger expressions. Here,
$\theta(\tau_M,...,\tau_1)\equiv\theta(\tau_M-\tau_{M-1})...\theta
(\tau_2-\tau_1)$.

Let us now go to the full theory by (functionally) integrating over the
gauge potentials and treating the interaction terms in the Yang-Mills
Lagrangian as perturbations. Correlators, with respect to the quadratic
terms in the action, of a given $A_\mu(x(\tau_m))$ entering (5) with
another gauge field produce gluonic propagators attached to fermionic
lines. In our specific case,
pairing two $A_\mu(x(\tau_m))$'s leads to a situation where a gluon is
emitted and subsequently absorbed by the fermionic line. Pairing with an
external field operator produces an attached external gluon. Finally, a
correlator with a field entering an interaction term in the gluonic
Lagrangian density connects the attachement with a three or four point
vertex\footnote{It is important to keep in mind that the fermionic line
has a space-time and {\it not} a Feynman diagrammatic interpretation.}.
Typical examples are displayed in Fig. 1. 

We isolate the following generic expression, after transcribing to
momentum representation for the gauge fields,
\begin{eqnarray}
i\Delta^{(M)}(x,y|A)&=&(ig)^M \int_0^\infty
dTe^{-Tm^2}\left(\prod_{n=M}^1\int_0^T
d\tau_n\right)\theta(\tau_M,...,\tau_1)\int_{\stackrel{x(0)=x}{x(T)=y}}
{\cal D}x(\tau)
[m-{1\over 2}\gamma\cdot\dot{x}(T)] \nonumber\\
& &\times  P\,exp\left[{i\over 4}\int_0^Td\tau\,\sigma\cdot
\omega\right]\tilde{A}(k_M)\cdot\dot{x}(\tau_M)...
\tilde{A}(k_1)\cdot\dot{x}(\tau_1)\nonumber\\ &&
\quad\times exp\left[-{1\over 4}\int_0^T d\tau {\dot{x}^2(\tau)}
-i\sum_{n=1}^M k_n\cdot x(\tau_n)\right],
\end{eqnarray}
which makes its appearance once we expand the Wilson line.

In order to
achieve direct comparisons with the Feynman diagram perturbative logic we
must first execute the path integral while rendering the spin
factor independent from the details of the specific path. This can be
achieved by employing the functional derivative operator [13,14] 
\begin{equation}
\frac{\delta}{\delta s_{\mu\nu}(t)}\equiv lim_
{\varepsilon
\rightarrow 0}\int^{+\varepsilon}_{-\varepsilon}d\sigma\, \sigma
\frac{\delta^2}{\delta x_\mu(t+\sigma/2)\delta x_\nu(t-\sigma/2)}.
\end{equation}
One easily determines [11]
\begin{equation}
\frac{\delta}{\delta s_{\mu\nu}(t)}exp\left[-{1\over 4}\int_0^Td\tau
\dot{x}^2(\tau)\right]={1\over 2}\omega_{\mu\nu}(x(t))\left[-{1\over 4}
\int_0^Td\tau \dot{x}^2(\tau)\right].
\end{equation}

Introducing the anticommuting $\xi$ variables by $\tilde{A}_n\cdot\dot{x}
(\tau_n)=\int d\xi_nd\bar{\xi}_ne^{\bar{\xi}_n\tilde{A}_n\cdot\dot{x}
(\tau_n)\xi_n}$  and setting
$\hat{k}_\mu(\tau_{n\mu})\equiv k_{n\mu}+
+g\bar{\xi}_n\xi_n 
\tilde{A}_{n\mu}\frac{\partial}{\partial\tau_n}$ we obtain

\begin{eqnarray}
i\Delta^{(M)}(x,y|A)&&=(ig)^M \int_0^\infty
dTe^{-Tm^2}\left(\prod_{n=M}^1\int_0^Td\tau_n\right)
\theta(\tau_M,...,\tau_1) \left(\prod_{n=M}^1\int d\xi_n
d\bar{\xi}_n\right) 
\nonumber\\& &\times\int_{\stackrel{x(0)=x}{x(T)=y}}{\cal D}x(\tau)
[m-{1\over 2}\gamma\cdot\dot{x}(T)]  e^{\left[-{1\over 4}\int_0^T d\tau
{\dot{x}^2(\tau)}
-i\sum_{n=1}^M \hat{k}_n\cdot x(\tau_n)\right]}\Phi[\hat{k}],
\end{eqnarray}
where
\begin{eqnarray}
\Phi[\hat{k}]=&& exp\left[i\sum_{n=1}^M \hat{k}_n\cdot x(\tau_n)\right]
\nonumber\\&&\times Pexp\left({i\over 2}\int_0^Tds\sigma_{\mu\nu}
\frac{\delta}{\delta s_{\mu\nu}}\right)
exp\left[-i\sum_{i=1}^M\hat{k}(\tau_n)\cdot x(\tau_n)\right]
\end{eqnarray}
is the spin-factor expression in the ``$\hat{k}$-representation".

The integral over paths can now be easily performed as the particle-like
action functional $S[x]={1\over 4}\int_0^T d\tau{\dot{x}^2(\tau)}
-i\sum_{n=1}^M \hat{k}_n\cdot x(\tau_n)$ does not contain non-linear
terms. The procedure has been deliberated in Ref. [11] and will be
carried out, in this paper, for the more complex `particle action
functionals' that will arise in section 4. Here, we give the final result 
which reads, in Fourier conjugate space, 
\begin{eqnarray}
i\tilde{\Delta}^{(M)}(p,p'|A)&=&(2\pi)^4(ig)^M\delta\left(p+p'- 
\sum_{n=1}^Mk_n\right) 
\int_0^\infty dTe^{-Tm^2}\left(\prod_{n=M}^1\int_0^T
d\tau_n\right)\theta(\tau_M,...,\tau_1)\nonumber\\&&\times 
\left(\prod_{n=M}^1\int d\xi_nd\bar{\xi}_n\right)[m-i\gamma\cdot p-
ig\sum_{n=1}^M\bar{\xi}_n\xi_n\gamma\cdot\tilde{A}_n\delta(\tau_n-T)]
\nonumber\\&&\times exp\left[-\sum_{n=1}^{M+1}\sum_{m=1}^{M+1}\hat{k}
(\tau_n)\cdot\hat{k}(\tau_m){\rm min}(\tau_n,\tau_m)\right]\Phi[\hat{k}]
\end{eqnarray} 
with $k_{M+1,\mu}\equiv p'_\mu,\,\tau_{M+1}\equiv T$ and $\tilde{A}_{M+1,
\mu}\equiv 0_\mu$.

The spin factor expression assumes a form, for a given $M$, which
explicitly displays its path independence. It reads as follows
\begin{eqnarray}
\Phi[\hat{k}]&=&Pexp\left(-ig\sum_{n=1}^M\sigma_{\mu_n\nu_n} 
f_{\mu_n\nu_n}(\hat{k})
\right)\nonumber\\&\stackrel{def}{=}&
1+(-ig)\sum_{n=1}^M\sigma_{\mu_n\nu_n}
f_{\mu_n\nu_n}+(-ig)^2\sum_{n=1}^M\sum_{m=1}^{n-1}\sigma_{\mu_n\nu_n}
\sigma_{\mu_m\nu_m}f_{\mu_n\nu_n}f_{\mu_m\nu_m}+...\, ,
\end{eqnarray}
where\footnote{It is evident that, due to the presence of the
$\xi$-variables, the sums entering (12) have, at most, $M$ terms. One
should also take note of the fact that the subscript $n$ implicitly
incorporates group indices.} 
\begin{eqnarray}
f_{\mu_m\nu_n}&\equiv& \bar{\xi}_n
\xi_n(\tilde{A}_{n\mu_m}k_{n\nu_n}
-\tilde{A}_{n\nu_n}k_{n\mu_m})\nonumber\\&&-ig\bar{\xi}_{n+1} 
\xi_{n+1}\bar{\xi}_n\xi_n(\tilde{A}_{n\mu_m}k_{n+1,\nu_n}-\mu
\leftrightarrow
\nu)\delta(\tau_{n+1}-\tau_n).
\end{eqnarray}
It can be easily verified that $\Phi[\hat{k}]$ provides the numerator
resulting from the 
string of fermionic propagators for $M$ gluonic insertions
on the the quark (world)line, with the corresponding denominators cast in
the form $p^2-m^2+i\epsilon$.

One concludes that, according to the Fock-Feynman-Schwinger path integral
casting of the field theory, a perturbative computation involving gauge
field attachments to a given fermionic line\footnote{Actually, if more
than one fermionic lines are involved, e.g. quark-quark ``scattering"
process, the total number of attachments to all lines should be counted.}
can be
organized according to the number $M$ of points where the gauge fields
``strike" the spin-1/2 particle (world)line(s). For a given Green's
function and to a given perturbative order the aformentioned number $M$
takes more than one value. On the other hand, for fixed $M$ the parametric
integrations should take care of all possible rearrangements, including
those of colour indices, of the struck points along the contour. An
example, corresponding to a ``Compton gluon scattering process",
will occupy our attention next.

\vspace{.2cm}

{\bf 3. A low order analytic computation.} Consider the QCD ``Compton"
process depicted in Fig. 2. At second perturbative order we have to
contend with an $M=2$ and an $M=1$ contribution.

For $M=2$ and given that for incoming/outgoing gluons $\epsilon_i\cdot
k_i=0$, we have
\begin{eqnarray}
\Phi[k]&=&1-ig\bar{\xi}_1\xi_1\gamma\cdot\epsilon_1\gamma\cdot
k_1-ig\bar{\xi}_2\xi_2\gamma\cdot\epsilon_2\gamma \cdot k_2
\nonumber\\&&-g^2\bar{\xi}_2\xi_2\bar{\xi}_1\xi_1
[\gamma\cdot\epsilon_2,\gamma\cdot\epsilon_1]\delta(\tau_2-\tau_1).
\end{eqnarray}

Substituting into the expression for $\Delta^{(2)}$, allowing for
momentum conservation, choosing $\epsilon_i\cdot p=0$ and performing the
integrations over the Grassmann variables we obtain
\begin{eqnarray}
i\tilde{\Delta}^{(2)}(p,p'|\{\epsilon\})&=&
-g^2(2\pi)^4\delta(p+p'+k_1+k_2)
\int_0^\infty dTe^{-T(m^2+p'^2)}\int_0^Td\tau_2\int_0^Td\tau_1
\theta(\tau_2-\tau_1)\nonumber\\&&\times
exp[-T(m^2+p'^2)-2p'\cdot k_2(\tau_2-\tau_1)]
\{(m-i\gamma\cdot p')\nonumber\\&&\times
[2\gamma\cdot\epsilon_1\gamma\cdot\epsilon_2\delta(\tau_1-\tau_2)
+2\epsilon_2\cdot p'\gamma\cdot k_1\gamma\cdot\epsilon_1
-\gamma\cdot k_1\gamma\cdot\epsilon_1 
\gamma\cdot k_2\gamma\cdot\epsilon_2\nonumber\\&&
+i\gamma\cdot\epsilon_2\gamma\cdot
k_1\gamma\cdot\epsilon_1\delta(\tau_2-T)\}.
\end{eqnarray}

Standard manipulations involving the gamma matrices along with the
restoration of (basically group) factors, omitted in $\Delta^{(2)}$,
easily lead to the following $M=2$ contribution to the amplitude 
\begin{eqnarray}
A^{(2)a_1a_2}_{Comp}&=&g^2(2\pi)^4\delta(p+p'+k_1+k_2)
\frac{1}{m+i\gamma\cdot p'}
\nonumber\\&&\times\left[T^{a_2}\frac{\gamma\cdot\epsilon_2
\gamma\cdot k_1\gamma\cdot\epsilon_1}{2p'\cdot k_2}T^{a_1}+
T^{a_1}\frac{\gamma\cdot\epsilon_1
\gamma\cdot k_2\gamma\cdot\epsilon_2}{2p'\cdot k_1}T^{a_2}\right]
\frac{1}{m-i\gamma\cdot p}.
\end{eqnarray}

For the $M=1$ contribution we have
\begin{eqnarray}
\Phi[k]&=&Pexp[-{1\over 2}g\bar{\xi}\xi\sigma_{\mu\nu}(\tilde{A}(k)_\mu
k_\nu-\tilde{A}(k)_\nu k_\mu)]\nonumber\\&=& 1-{1\over 2}g\bar{\xi}\xi
[\gamma\cdot\tilde{A}(k),\gamma\cdot k]
\end{eqnarray}
(note that this time the attached gauge field is internal) leading to
\begin{eqnarray}
i\tilde{\Delta}^{(1)}(p,p'|\tilde{A})&=& -g(2\pi)^4\delta(p+p'+k_1+k_2)
\int_0^\infty dTe^{-T(m^2+p'^2)}\int_0^Td\tau\{
(m-i\gamma\cdot p')
\nonumber\\&&\times((p'-p)\cdot\tilde{A}(k)+{1\over 2}
[\gamma\cdot\tilde{A}(k),\gamma\cdot k])+i\gamma\cdot\tilde{A}(k)
\delta(\tau-T)\}
\end{eqnarray}
which, after some gamma-matrix algebra routine and once taking into
account the presence of free spinor wavefunctions at each end of the
fermionic line, gives
\begin{equation}
i\tilde{\Delta}^{(1)}(p,p'|\{k\})=-ig 
\frac{1}{m+i\gamma\cdot p'} \gamma\cdot\tilde{A}(k)\frac{1}{m-i\gamma\cdot
p}.
\end{equation}

The remaining work leading to the amplitude contribution involves standard
perturbation theory manipulations related to the `merging' of the attached
gluon with the three point Yang-Mills vertex. It  now becomes imperative
to
restore the color indices for the gauge fields, so let us assign index $a$
to the gluon coming from the Wilson line, corresponding to the attachment
on the fermionic contour, and indices $b,\,c,\,d$ to the vertex gluons.
Clearly, two of the latter register as incoming/outgoing states hence they
are associated with polarization vectors. 

In all, the final $M=1$ contribution to the amplitude is 
\begin{eqnarray}
A^{(1)a_1a_2}_{Comp}&=&(2\pi)^4\delta(p+p'+k_1+k_2)g^2T^cf^{ca_1a_2}
\frac{1}{m+i\gamma\cdot p'}\gamma_\rho\epsilon_{1\mu}\epsilon_{2\nu} 
\nonumber\\&&\times\left[\delta_{\mu\nu}(k_2-k_1)_\rho-
\delta_{\rho\nu}(2k_2+k_1)_\mu+\delta_{\mu\rho}(k_2+2k_1)_\nu\right]
\frac{1}{(k_1+k_2)^2}\frac{1}{m-i\gamma\cdot p}
\end{eqnarray}
which, combined with equ (16), yields the
correct final result.

A few comments are in order at this point. First, the present example
serves more as an illustration of the fact that the Fock-Feynman-Schwinger
recasting of QCD, aided by the presence of the spin factor, provides a
viable methodology
for conducting perturbative calculations. The real test concerning
possible advantages of this approach should come with respect to higher
order calculations where a plethora of Feynman diagrams enters. Efforts
are under way to probe such a prospect via the employment of numerical
methodologies. Second, the fact that purely gluonic perturbative
contributions have been treated conventionally in the above application
does not preclude the possibility of furnishing the gluonic sector with
a worldline description. Of special notice, in this respect, is the work
in Ref [15] which, even though conducted in the string context,
adopts a space-time mode of description while addressing itself to the
gauge
field sector. An alternative possibility within the framework of
our approach  {\it is} viable and will be presented next. Its basic
feature is that, as a perturbative scheme, it deviates from the Feynman
diagram logic by sidesteping the momentum representation and arriving
at final expressions which solely involve (one-dimensional) path
integrations.

\vspace{.2cm}

{\bf 4. A space-time approach to perturbative expansions.} We shall now
switch policy and leave the task of carrying out the path integral until
after we have explicitly dealt with the gauge field 
sector. Insisting on the perturbative treatment let us consider a given
configuration entering a process of interest contributing
to some fixed order. Introducing for each pairing of gauge fields the
corresponding space-time correlator, we obtain expressions which no longer
involve vector potentials.

Now, the general form of a gauge field correlator in
configuration space, in the Feynman gauge, is
\begin{equation}
<A_\mu^a(x)\,A_\nu^b(y)>_0=\delta^{ab}\delta_{\mu\nu}
\frac{\Gamma\left({D\over 2}-1\right)}{4\pi^{D/2}|x-y|^{D -2}}.
\end{equation}
Clearly, if $x=x(\tau)$ and/or $y=y(\tau)$, then one, or both, arguments
of the gauge fields lie on a fermionic particle's contour, hence the
corresponding correlator is directly fed into the path integral. Under
these circumstances the latter cannot be performed in the same manner as
before. One might say that the induced `disappearance' of the gauge fields
from our expressions calls for a readjustment on how the separation
between geometry and dynamics is to be effected. Our new strategy will be
to incorporate both the spin factor and the gauge field correlator 
into an overall expression that will define a
`particle action functional' with respect to which velocity-velocity
correlators are to be computed. The latter arise as remnants of the
expansion of the Wilson exponential(s), once the gauge fields have been
removed. 

In order to form the aformentioned action functional it becomes
imperative to get rid of the second derivatives entering the expression
for $\omega_{\mu\nu}(x)$. To confront this task we must first devise
an alternative way
of displaying the $\gamma$-matrix path ordering in the spin factor.
To this end we follow Ref [16] and introduce auxiliary anticommuting
variables $\psi_\mu(\tau)$ obeying the relations
$\{\psi_\mu(\tau),\psi_\nu(\tau')\}=0$ and
$\{\psi_\mu(\tau),\psi_\nu(\tau)\}={1\over 2}\delta_{\mu\nu}$. One can
then show that 
\begin{equation}
trPexp\left\{\frac{i}{4}\int_0^Td\tau\,\sigma\cdot\omega\right\}= {1\over
D}\int_{\psi_\mu(0)+\psi_\mu(T)=0}[d\psi] exp\left[
\int_0^Td\tau\psi(\tau)\cdot\dot{\psi}(\tau)-\int_0^Td\tau\psi_\mu\psi_\nu
\omega_{\mu\nu}\right]
\end{equation}
relevant for closed and
\begin{eqnarray}
Pexp\left\{\frac{i}{4}\int_0^Td\tau\,\sigma\cdot\omega\right\} 
&=&exp\left(i\gamma\cdot\frac{\partial}{\partial\lambda}
\right)\int_{\psi_\mu(0)+\psi_\mu(T)=\lambda_\mu}[d\psi] exp\left[
\int_0^Td\tau\psi(\tau)\cdot\dot{\psi}(\tau)\right.\nonumber\\ &&\left.
-\int_0^Td\tau \psi_\mu\psi_\nu\omega_{\mu\nu} 
+\psi(T)\cdot\psi(0)\right]_{\lambda=0}
\end{eqnarray} 
relevant for open particle contours, respectively. 

We shall look upon the factor
$exp\left[\int_0^Td\tau\psi(\tau)\cdot\dot{\psi}(\tau)\right]$ entering
(23) as furnishing a fermionic sector to the `particle action' thereby
formulating a `super particle' description. For future reference we note
that
\begin{eqnarray}
<\psi_\mu(t)\psi_\nu(t')>_{\psi,\lambda}&&\equiv
\int_{\psi_\mu(0)+\psi_\mu(T)=\lambda_\mu}[d\psi] exp\left[
\int_0^Td\tau\psi(\tau)\cdot\dot{\psi}(\tau) +\psi(0)\cdot\psi(T)\right]
\psi_\mu(t)\psi_\nu(t')
\nonumber\\&&\quad =-{1\over 4}\delta_{\mu\nu} sign(t-t')\lambda_\mu
\lambda_\nu
\end{eqnarray}
with $sign\,0=0$.

The next step is to write
\begin{equation}
\int_0^Td\tau \psi_\mu\psi_\nu\omega_{\mu\nu}=-T\int_0^Td\tau
(\psi\cdot\dot{x})^\cdot(\psi\cdot\dot{x})+T\int_0^Td\tau
(\psi\cdot\dot{x})(\dot{\psi}\cdot\dot{x})+{T\over 4}\int_0^Td\tau
\dot{x}\cdot\ddot{x}
\end{equation} 
whose usefulness will become evident once we specify the correlators
$<\dot{x}_\mu(t)
\dot{x}_\nu(t')>_x$ and $<\psi_\mu(t)\dot{\psi}_\nu(t')>_\psi$ in the
limit where the two arguments coincide. To this end we define
\begin{eqnarray}
<\dot{x}_\mu(t)\dot{x}_\nu(t)>_x&\stackrel{def}{=}&-
lim_{\epsilon\rightarrow 0}{1\over T}
\int_{-\epsilon}^{+\epsilon}d\sigma\,\frac{\delta^2}{\delta J_\mu (t+
{\sigma\over 2}) \delta J_\nu (t-{\sigma\over 2})}\nonumber\\ &&
\quad\times<e^{i\int_0^Td\tau J(\tau)\cdot\dot{x}(\tau)}>_x|_{J=0}
={2\over T}\delta_{\mu\nu},
\end{eqnarray}
reflecting the fact that we are integrating over paths with
$|\dot{x}|=const$ and,
similarly, ($\eta_\mu$ and $\theta_\mu$ anticommuting sources)
\begin{eqnarray}
<\psi_\mu(t)\dot{\psi}_\nu(t)>_\psi&\stackrel{def}{=}&-
lim_{\epsilon\rightarrow 0}{1\over T}
\int_{-\epsilon}^{+\epsilon}d\sigma\,\frac{\delta^2}{\delta \eta_\mu
(\tau+
{\sigma\over 2}) \delta \theta_\nu (\tau-{\sigma\over 2})}\nonumber\\ && 
\times<e^{i\int_0^Td\tau(\eta\cdot\psi+\theta\cdot\dot{\psi} 
)}>_\psi|_{\eta=\theta=0} ={1\over 2T}\delta_{\mu\nu}
\end{eqnarray}
whereupon we are led to the substitution 
rules\footnote{Note that only the $\int_0^Td\tau\dot{x}^2$ part of the
bosonic particle action contributes to (26) so any additional terms are
imatterial as far as our definition of velocity-velocity correlators
at equal times is concerned.}
$\dot{x}_\mu(t)\dot{x}_\nu(t)\rightarrow{2\over T}\delta_{\mu\nu},
\,\psi_\mu(t)\dot{\psi}_\nu(t)\rightarrow{1\over 2T}\delta_{\mu\nu}$. It,
thereby, follows that (25) assumes the form
\begin{equation}
\int_0^Td\tau \psi_\mu\psi_\nu\omega_{\mu\nu}=-T\int_0^Td\tau
(\psi\cdot\dot{x})(\psi\cdot\dot{x})^\cdot+const.
\end{equation}
Designating ${1\over \sqrt{T}}\psi_5(\tau)\equiv\psi(\tau)\cdot
\dot{x}(\tau)$ and enforcing this definition through a delta-function
costraint (enter Grassmann path variable $\chi(\tau)$) we
find\footnote{From hereon we restrict our attention
to open line expressions.}
\begin{eqnarray}
Pexp\left\{\frac{i}{4}\int_0^Td\tau\,\sigma\cdot\omega\right\}
&=&exp\left(i\gamma\cdot\frac{\partial}{\partial\lambda}
\right)\int_{\psi_\mu(0)+\psi_\mu(T)=\lambda_\mu}[d\psi(\tau)]
\int[d\psi_5(\tau)]\int[d\chi(\tau)]\nonumber\\&&\times
exp\left\{\int_0^Td\tau\psi\cdot\dot{\psi}
+\int_0^Td\tau\psi_5\dot{\psi}_5+{i\over \sqrt{T}}
\int_0^Td\tau \chi(\tau)\psi_5(\tau)\right.\nonumber\\&& 
-i\int_0^Td\tau \chi(\tau)\psi(\tau)
\cdot\dot{x}+\psi(T)\cdot\psi(0)\left.\right\}_{\lambda=0}.
\end{eqnarray}

For the purpose of computing velocity-velocity correlators we need
to extract
from the expression on the right hand side that part which is relevant
to forming the `bosonic sector' of the `particle action'. This is
furnished by the term $-i\int_0^T 
d\tau{\cal J}\cdot\dot{x}$, where
\begin{equation}
{\cal J}_\mu(\tau)\equiv \chi(\tau)\psi_\mu(\tau).
\end{equation}

A further contribution to the `bosonic particle action' comes from the
propagator 
expression. Exponentiating the denominator entering (21), via the 
employment of (Feynman) auxiliary integration variables $\alpha_i$ we
obtain 
\begin{equation}
S_2[x]={1\over 4}\int_0^Td\tau\dot{x}^2+i\int_0^Td\tau {\cal J}(\tau)
\cdot \dot{x}(\tau)+\sum_{i=1}^M\alpha_i[x(\tau_{a_i})-x(\tau_{b_i})]^2
\end{equation}
representing all double-ended gauge field attachments and
\begin{equation}
S_1[x]={1\over 4}\int_0^Td\tau\dot{x}^2+i\int_0^Td\tau {\cal J}(\tau)
\cdot \dot{x}(\tau)+\sum_{i=1}^M\alpha_i[x(\tau_i)-z_i]^2
\end{equation}
for single-end attachments to the matter field contour. 

Clearly, unattached gauge field propagators are `averaged' with a
`particle action' that does not contain the third term. Moreover, one
should not forget that there are fermionic path integrals which need to be
performed. In what follows we shall focus our efforts on determining 
velocity-velocity correlators for each of the above two action
functionals. 

Consider, first, the case of a single attachment, i.e. functional
$S_1[x]$. The `classical equations of motion' read
\begin{equation}
\ddot{x}^{cl}_\mu(\sigma)=-2i\dot{{\cal J}}_\mu(\sigma)
+4\sum_{i=1}^M\alpha_i[x(\tau_i)-z_i]\delta(\tau_i-\sigma).
\end{equation}
Making a variable change specified by $x_\mu(\sigma)=w_\mu(\sigma)
+\frac{(y-x)_\mu}{T}\sigma +x_\mu$, which obeys the, simpler, boundary 
conditions\footnote{Recall that the corresponding conditions 
for $x_\mu(\sigma)$ are
$x_\mu(0)=x_\mu$ and $x_\mu(T)=y_\mu$.} $w_\mu(0)=w_\mu(T)=0$ the
equations of motion assume the form
\begin{equation}
\ddot{w}^{cl}_\mu(\sigma)-b(\sigma){w}^{cl}_\mu(\sigma)=f_\mu(\sigma),
\end{equation}
where $f_\mu(\sigma)\equiv -2i\dot{{\cal J}}_\mu(\sigma)
+4\sum_{i=1}^M\alpha_i[\frac{(y-x)_\mu}{T}\tau_i +x_\mu-z_{i\mu}]
\delta(\tau_i-\sigma)$ and
$b(\sigma)\equiv 4\sum_{i=1}^M\alpha_i\delta(\tau_i-\sigma)$.

Introducing the Green's function $K^{(N)}(\sigma,\sigma')$ by
\begin{equation}
\left[\frac{d^2}{d\sigma^2}-b(\sigma)\right]K^{(N)}(\sigma,\sigma')=
\delta(\sigma-\sigma'),
\end{equation}
obeying boundary conditions $K^{(N)}(0,\sigma')=K^{(N)}(T,\sigma')=0$, we
write
\begin{equation}
w^{cl}_\mu(\sigma)=\int_0^T d\sigma'K^{(N)}(\sigma,\sigma')f_\mu(\sigma').
\end{equation}

Manipulations involved in the determination of the Green's function
$K^{(N)}(\sigma,\sigma')$ proceed along the lines employed in Ref
[17], adopted also in Ref [11], leading to the result
\begin{equation}
K^{(N)}(\sigma,\sigma')=-\Delta(\sigma,\sigma')+\sum_{i,j=1}^N
\hat{\Delta}(\sigma,\tau_i)\hat{\Delta}(\tau_i,\tau_j)
\hat{\Delta}(\tau_j,\sigma'),
\end{equation}
where
\begin{equation}
\Delta(\sigma,\sigma')=-<\sigma'|\partial^{-2}|\sigma>=
\frac{\sigma(T-\sigma')}{T}\theta(\sigma'-\sigma)+
\frac{\sigma'(T-\sigma)}{T}\theta(\sigma-\sigma')
\end{equation}
while the $\hat{\Delta}$'s are related to the $\Delta$'s through a
multiplication of the latter by a factor $\sqrt{\alpha_i}$ for each
$\tau_i$ in the argument. 

Consider, finally, the velocity-velocity correlator
\begin{eqnarray}
<\dot{x}_\mu(t_1)\dot{x}_\nu(t_2)>_{S_1}&=&-\frac{\delta^2}
{\delta{\cal J}_\mu(t_1)\delta{\cal J}_\nu(t_2)}\int_{x(0)=x,x(T)=y}
{\cal D}x(\tau)e^{-S_1[x]}\nonumber\\&=&-{\cal N}(\alpha)\frac{\delta^2}
{\delta{\cal J}_\mu(t_1)\delta{\cal J}_\nu(t_2)}exp\{-S_1[x]\},
\end{eqnarray}
where
\begin{equation}
{\cal N}(\alpha)=\int_{x(0)=x(T)=0}{\cal D}x(\tau)exp\left[
-{1\over 4}\int_0^Td\tau\dot{x}^2-\sum_{i=1}^N\alpha_ix^2(\tau_i)\right]
={1\over (4\pi T)^2}det^{-2N}(1+4\hat{\Delta}).
\end{equation}

Straight forward manipulations lead to the result
\begin{equation}
<\dot{x}_\mu(t_1)\dot{x}_\nu(t_2)>_{S_1}={\cal N}(\alpha)
\left[\dot{x}_\mu^{cl}(t_1)\dot{x}_\nu^{cl}(t_2)-2\delta_{\mu\nu}
\frac{\partial^2}{\partial t_1\partial t_2}K^{(N)}(t_1,t_2)\right]
e^{-S_1[x^{cl}]}
\end{equation}
with all the `classical' quantities readily calculable.

The case with $S_2[x]$, even though technically somewhat more complex,
proceeds along parallel lines. It becomes convenient to introduce [17] the
kernel
\begin{equation}
B(\sigma_1,\sigma_2)\equiv\sum_{i=1}^N\alpha_i[\delta(\sigma_1-\tau_{a_i})
-\delta(\sigma_1-\tau_{b_i})][\delta(\sigma_2-\tau_{a_i})
-\delta(\sigma_2-\tau_{b_i})]
\end{equation}
which allows us to cast the third term on the right hand side of equ (31)
in the form $\int_0^Td\sigma_1\int_0^Td\sigma_1B(\sigma_1,\sigma_2)
x(\sigma_1)\cdot x(\sigma_2)$. Accordingly, the relevant Green's function
$\Delta^{(N)}(\sigma,\sigma')$ now satisfies the equation
\begin{equation}
\int_0^Td\sigma_1\left[\delta(\sigma-\sigma_1)\frac{\partial^2}{\partial
\sigma_1^2}-4B(\sigma_1,\sigma)\right]\Delta^{(N)}(\sigma_1,\sigma')
=\delta(\sigma-\sigma'),
\end{equation}
where
\begin{equation}
\Delta^{(N)}(\sigma,\sigma')=-\Delta(\sigma,\sigma')+\sum_{i,j=1}^N
[\hat{\Delta}(\sigma,\tau_{a_i})-\hat{\Delta}(\sigma,\tau_{b_i})]
\hat{D}_{ij}[\hat{\Delta}(\sigma,\tau_{a_j})-\hat{\Delta}(\sigma,\tau_{b_j})]
\end{equation}
with
\begin{equation}
\hat{D}_{ij}\equiv \hat{\Delta}(\tau_{a_i},\tau_{a_j})
+\hat{\Delta}(\tau_{b_i},\tau_{b_j})
-\hat{\Delta}(\tau_{a_i},\tau_{b_j}) 
-\hat{\Delta}(\tau_{b_i},\tau_{a_j}). 
\end{equation}

The final
result for the velocity-velocity correlator acquires the corresponding
form
\begin{equation}
<\dot{x}_\mu(t_1)\dot{x}_\nu(t_2)>_{S_2}=\hat{{\cal N}}(\alpha)
\left[\dot{x}_\mu^{cl}(t_1)\dot{x}_\nu^{cl}(t_2)-2\delta_{\mu\nu}
\frac{\partial^2}{\partial t_1\partial t_2}\Delta^{(N)}(t_1,t_2)\right]
e^{-S_2[x^{cl}]},
\end{equation} 
where,
\begin{equation}
\hat{{\cal N}}(\alpha)={1\over (4\pi T)^2}det^{-2N}(1+4\hat{D}).
\end{equation}

Path integrals over the fermionic variables are much simpler, as can
be witnessed from the relevant manipulations performed in Ref [11]. The
basic result is displayed by equ (24).

\vspace{.2cm}

{\bf 5. Summary.} We have applied the Fock-Feynman-Schwinger path integral
to perturbative expansions in QCD for processes involving an open
fermionic line. The novel feature of the formalism is that it contains
a spin factor and a dynamical factor (Wilson line) whose distinguishable
roles facilitate the emergence of comprehensive expressions for the
perturbative series. Two
alternative approaches have been discussed. According to the first one the
path integral is performed right away while the spin factor is recast into
a momentum representation form which communicates with the gauge sector.
In this way one achieves direct comparison with the Feynman diagrammatic
expansion as the gauge fields enter through conventional momentum
space propagators while the spin factor expression provides the
contribution
coming from a fermionic line that has a given number of gauge field 
attachments. All this enters a master expression with parametric
integrations which account for all possible rearrangements of the points
on the fermionic line where the attachments occur. The second approach has
as its starting point the removal (in pairs) of the gauge fields in favor
of correlator expressions in a space-time representation. The spin factor,
on the other hand, is now trancribed into a form that contributes to
a superparticle-like (one-dimensional) action given that it involves
anticommuting coordinates as well. The result of this procedure is the
emergence of
a space-time picture for the perturbative series which boils down to
computing correlators of the type $<\dot{x}_\mu(t)
\dot{x}_\nu(t')>_x$ and $<\psi_\mu(t)\psi_\nu(t')>_\psi$
in a particle-based representation
of the theory (QCD). All that remains is to perform integrations over
the Feynman parameters. Non-trivial applications of the proposed scheme
are currently in pursuit.

\vspace{1cm}

{\bf Figure Captions}
\vspace{.2cm}

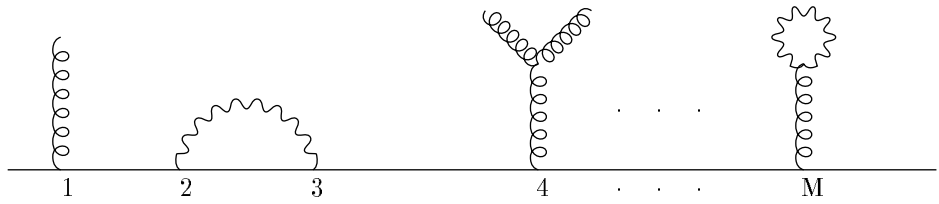
Fig. 1: Gluonic field attachments on a fermionic (world)line. The Wilson
line, which carries the dynamics between matter and gauge field quanta in
the Fock-Feynman-Schwinger path integral approach to QCD, is responsible
for the emergence of master expressions organized according to the number
M of such attachments. The spin factor, on the other hand, provides the
numerator of the string of fermionic propagators corresponding to a given
number of insertions.

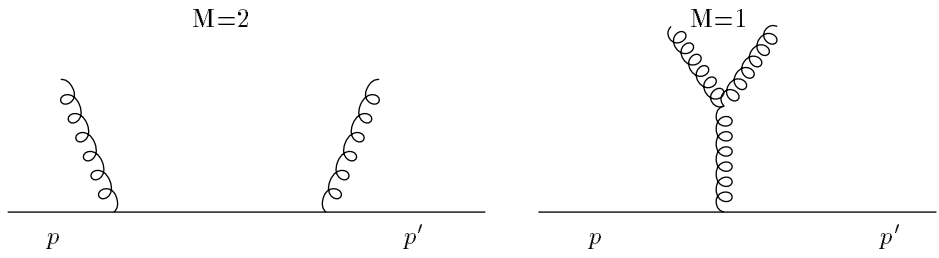
Fig. 2: Configurations entering the `QCD Compton Scattering amplitude', to
the lowest order, organized according to number of attachments onto the
fermionic (world)line.


\begin{thebibliography}{99} 
 
\bibitem{1} V. A. Fock, Izvestiya Akad. Nauk USSR, OMEN (1937) 557.

\bibitem{2} R. P. Feynman, Phys. Rev. {\bf 80}, (1950) 440.

\bibitem{3} J. Schwinger, Phys. Rev. {\bf 82}, (1951) 664.

\bibitem{4} E. S. Fradkin, Nucl. Phys. B {\bf 76} (1966) 588.

\bibitem{5} M. B. Halpern, A. Jevicki and P. Senjanovic, Phys. Rev. D {\bf
16} (1977) 2476; M. B. Halpern and W. Siegel, Phys. Rev D {\bf 16} (1977)
2486.

\bibitem{6} M. J. Strassler, Nucl. Phys. B {\bf 385} (1992) 145.

\bibitem{7} A. I. Karanikas and C. N. Ktorides, Phys. Lett B {\bf 275}
(1992) 403.

\bibitem{8} A. M. Polyakov, in {\it Fields, Strings and Critical
Phenomena}, edited by E. Br\'{e}zin and J. Zinn-Justin (North Holland,
Amsterdam, 1990). 

\bibitem{9} I. A. Korchemskaya and G. P. Korchemsky, J. Phys. A {\bf 24}
(1991) 4511; G. P. Korchemsky, Phys. Lett. B {\bf 257} (1993) 124. 

\bibitem{10} A. I. Karanikas and C. N. Ktorides, Phys. Rev. D {\bf 52}
(1995) 5883.   

\bibitem{11} A. I. Karanikas and C. N. Ktorides, JHEP {\bf 11} (1999) 033.

\bibitem{12} Z. Bern and D. A. Kosower, Phys. Rev. D {\bf 38}
(1988) 1888; Nucl. Phys. {\bf B 321} (1989) 451. 

\bibitem{13} A. M. Polyakov, Nucl. Phys. B {\bf 164} (1979) 171.

\bibitem{14} A. M. Migdal, Phys. Rep. C {\bf 102} (1983) 199.

\bibitem{15} P. Di Vecchia, L. Magnea, A. Lerda, R. Russo and R. Marotta,
Nucl. Phys. {\bf B 469} (1996) 235.

\bibitem{16} E. S. Fradkin and D. M. Gitman, Phys. Rev. D {\bf 44}
(1991) 3230.

\bibitem{17} M. G. Schmidt and C. Schubert, Phys. Lett. B {\bf 331} (1994)
69; {\it ibid.} Phys. Rev. D {\bf 53} (1996) 2150.
                                                  

\end{thebibliography}
\end{document}